\documentclass[sn-nature]{sn-jnl}

\usepackage{graphicx}%
\usepackage{multirow}%
\usepackage{amsmath,amssymb,amsfonts}%
\usepackage{amsthm}%
\usepackage{mathrsfs}%
\usepackage[title]{appendix}%
\usepackage{xcolor}%
\usepackage{textcomp}%
\usepackage{manyfoot}%
\usepackage{booktabs}%
\usepackage{algorithm}%
\usepackage{algorithmicx}%
\usepackage{algpseudocode}%
\usepackage{listings}%

\usepackage[caption=false,font=normalsize,labelfont=sf,textfont=sf]{subfig}

\begin{document}

\title[Article Title]{Mixed Precision Photonic Computing with 3D Electronic-Photonic Integrated Circuits }


\author*[1]{\fnm{Georgios} \sur{Charalampous}}\email{gcharalampous@ucdavis.edu}

\author[2]{\fnm{Rui} \sur{Chen}}\email{charey@uw.edu}

\author[1]{\fnm{Mehmet} \sur{Berkay On}}\email{mbon@ucdavis.edu}

\author[4]{\fnm{Aslan} \sur{Nasirov}}\email{aslan.nasirov@vanderbilt.edu}

\author[5]{\fnm{Chun-Yi} \sur{Cheng}}\email{briancheng831@tamu.edu}

\author[1]{\fnm{Mahmoud} \sur{AbdelGhany}}\email{mabdelghany@ucdavis.edu}

\author[2,3]{\fnm{Arka} \sur{Majumdar}}\email{arka@uw.edu}

\author[1]{\fnm{Ji} \sur{Wang}}\email{jiiwang@ucdavis.edu}

\author[6]{\fnm{Jennifer A.} \sur{Black}}\email{jennifer.black@nist.gov}

\author[7]{\fnm{Rajkumar Chinnakonda} \sur{Kubendran}}\email{rajkumar.ece@pitt.edu}

\author[4]{\fnm{Caglar} \sur{Oskay}}\email{caglar.oskay@vanderbilt.edu}

\author[1]{\fnm{Zhaojun} \sur{Bai}}\email{zbai@ucdavis.edu}

\author[5]{\fnm{Sam} \sur{Palermo}}\email{spalermo@tamu.edu}

\author[6]{\fnm{Scott B.} \sur{Papp}}\email{scott.papp@nist.gov}

\author*[1]{\fnm{and S. J. Ben} \sur{Yoo}}\email{sbyoo@ucdavis.edu}

\affil*[1]{\orgdiv{Department of Electrical and Computer Engineering}, \orgname{University of California Davis}, \orgaddress{\street{One Shields Avenue}, \city{Davis}, \postcode{CA 95616}, \country{USA}}}

\affil[2]{\orgdiv{Department of Electrical and Computer Engineering}, \orgname{University of Washington}, \orgaddress{\street{185 Stevens Way Paul Allen Center}, \city{Seattle}, \postcode{WA 98195-2500}, \country{USA}}}

\affil[3]{\orgdiv{Department of Physics}, \orgname{University of Washington}, \orgaddress{\street{185 Stevens Way Paul Allen Center}, \city{Seattle}, \postcode{WA 98195-2500}, \country{USA}}}

\affil[4]{\orgdiv{Department of Civil and Environmental Engineering}, \orgname{Vanderbilt University}, \orgaddress{\street{2301 Vanderbilt Place}, \city{Nashville}, \postcode{TN 37235-1831}, \country{USA}}}

\affil[5]{\orgdiv{Department of Electrical and Computer Engineering}, \orgname{Texas A\&M University}, \orgaddress{\street{Wisenbaker Engineering Building 3128, 188 Bizzell St, College Station}, \city{Texas}, \postcode{TX 77843}, \country{USA}}}

\affil[6]{\orgdiv{Time and Frequency Division}, \orgname{National Institute of Standards and Technology}, \orgaddress{\city{Boulder}, \postcode{CO}, \country{USA}}}

\affil[7]{\orgdiv{Department of Electrical and Computer Engineering}, \orgname{ University of Pittsburgh}, \orgaddress{\street{1140 Benedum Hall}, \city{Pittsburgh}, \postcode{PA 15261}, \country{USA}}}


\abstract{We propose advancing photonic in-memory computing through 3D-Photonic-Electronic integrated circuits using Phase-Change-Material (PCM), and AlGaAs-CMOS technology. These circuits offer precision (\textgreater12-bits), scalability (\textgreater1024\texttimes1024), and parallelism (\textgreater1 million) in wavelength-space-time domains at ultra-low power (\textless1 W/PetaOPS). Monolithically integrated hybrid PCM AlGaAs memory resonators handle coarse-precision iterations (\textgreater 5-bit MSB precision) through phase-transitions in PCM. Electro-optic memristive tuning ensures high-precision iteration (\textgreater 8-bit LSB precision) for over 12-bits precision in-memory computing. PCM material with low loss (\textless 0.01 dB/cm) and electro-optical tuning yield memristive optical resonators with a high Q-factor (\textgreater10\textsuperscript{6}), low-loss, and low-power-tuning. The crossbar photonic tensor core, with W~\texttimes~W PCM AlGaAs memresonators, enables a general matrix multiply (GEMM) system for W wavelengths from optical frequency combs with low loss and minimal crosstalk. Hierarchical scaling of the W~\texttimes~W photonic tensor core in the wavelength domain (K) and spatial domain (L) addresses high-dimensional (N) scientific Partial Differential Equation (PDE) problems in a single operation O(1), contrasting with conventional O(N\textsuperscript{2}) complexity.}




\maketitle

\section{Introduction}
	Traditional computers follow a centralized processing architecture, characterized by a central processor and segregated memory, tailored for executing sequential, digital, procedure-based programs. Although the von-neuman architecture is generalized and flexible, it proves inefficient for computational models requiring distribution, massive parallelism, and adaptability, particularly those employed in matrix multiplications such as neural networks, iterative optimization algorithms, and partial differential equation (PDE) solvers~\cite{Shastri:21}.
	
	Optical computing is a paradigm of computation that utilizes the principles of optics, specifically the properties of light, to perform various computational tasks. Unlike traditional electronic computing, which relies on electrical signals to represent and process information, optical computing leverages photons (light particles) to carry and manipulate data. Photons travel at the speed of light, which is much faster than the speed of electrons in traditional electronic circuits. This high-speed property of light enables rapid data transmission and processing. Light waves can be manipulated in parallel, allowing for the simultaneous processing of multiple pieces of information. This inherent parallelism holds the potential for significantly faster computations in certain applications. 

	The goal of photonic processors should not be to replace conventional computers, but to enable applications that are unreachable at present by conventional computing technology—those requiring low latency, high bandwidth and low energies such as in communication networks, medical imaging, machine learning and artificial intelligence, security and encryption~\cite{Paul:17}.

	Addressing partial differential equations (PDEs) through numerical methods frequently demands extensive computational time, significant energy expenditure, and substantial hardware resources in real-world applications. Consequently, the widespread application of PDE solutions is constrained in various scenarios, such as autonomous systems and supersonic flows, where there is a constrained energy budget and a necessity for nearly instantaneous responses.
	
	As an illustration, consider the critical role of solving Hamiltonian-Jacobi-Issac (HJI) PDEs or Hamiltonian-Jacobi-Bellman (HJB) PDEs in the safety verification and control of autonomous systems. These equations need to be solved iteratively as sensor data evolves and avoidance specifications are updated. Unfortunately, training a Physics-Informed Neural Network (PINN) on a high-performance GPU can be a time-intensive process, requiring more than 20 hours~\cite{Onken:21,Somil:20}. This prolonged computational time poses a significant challenge, especially when there are stringent demands on the latency and energy cost of embedded computing platforms crucial for the operation of autonomous systems. Consequently, this impediment hinders the realization of real-time safety-aware decision-making capabilities in autonomous systems. Addressing this challenge is vital to enhance the efficiency and responsiveness of autonomous systems in dynamic environments.
	
	Accelerators based on Optical Neural Networks (ONN) show great potential for real-time inference and training \cite{Shen:17,Feldmann:21}. Nevertheless, the training of PINNs on photonic chips faces significant challenges, primarily due to three constraints. To begin with, photonic multiply-accumulate (MAC) units, such as Mach-Zehnder interferometers (MZIs), exhibit a significantly larger size on the order of tens of microns compared to CMOS transistors, leading to lower integration density. A PINN of actual size, featuring over 10\textsuperscript{5} model parameters, can readily surpass the available chip size according to the square scaling rule. In this rule, an N~\texttimes~N optical weight matrix necessitates O(N\textsuperscript{2}) Mach-Zehnder interferometers (MZIs)~\cite{Reck:94,Wetzstein:20}. Secondly, achieving on-chip training on photonic chips presents a challenge. Various back-propagation (BP)-free methods have been proposed to address the 'hardware-unfriendly' nature of error feedback in traditional back-propagation. Unfortunately, these methods are also limited by their scalability issue. Thirdly, the loss incurred during PINN training involves higher-order derivatives, necessitating multiple backpropagations (BPs) for accurate computation. Given the inefficiency of in-situ backpropagation \cite{Hughes:20}, an alternative numerical method is essential for the photonic implementation. Finally, the loss incurred during PINN training involves large number of iterations for accurate computation and convergence.

	Our proposed architecture work is dedicated to the realization of photonic in-memory computing through the integration of 3D-Photonic-Electronic circuits, incorporating Phase-Change-Material (PCM), AlGaAs, and CMOS technologies. The primary goals include achieving an exceptional level of accuracy surpassing 12-bits, ensuring high scalability exceeding 1024~\texttimes~1024 array dimensions, and implementing extreme parallelism within the wavelength-space-time domains, surpassing a remarkable 1~million parallel processes. All of this is to be achieved at an ultra-low power consumption of less than 1~Watt per PetaOPS.
	
	This work involves comprehensive design, simulation, validation, and bench-marking of a groundbreaking modality of scalable, ultra-low power 'in-memory' computation. This novel approach is characterized by its exceptionally low Size, Weight, and Power (SWaP) requirements, promising high throughput, and adaptive programmability. The anticipated outcomes of this research hold the potential to revolutionize computing paradigms, offering a versatile solution applicable across a wide spectrum of applications. The focus lies not only on pushing the boundaries of computational accuracy and scalability but also on ensuring efficiency and adaptability in real-world scenarios. Through this innovative approach, we aim to usher in a new era of computing that aligns with the demands of various applications while operating at the forefront of technological advancements.

	\begin{figure}[tb!]
		\centering
		\includegraphics[width=1.0\textwidth]{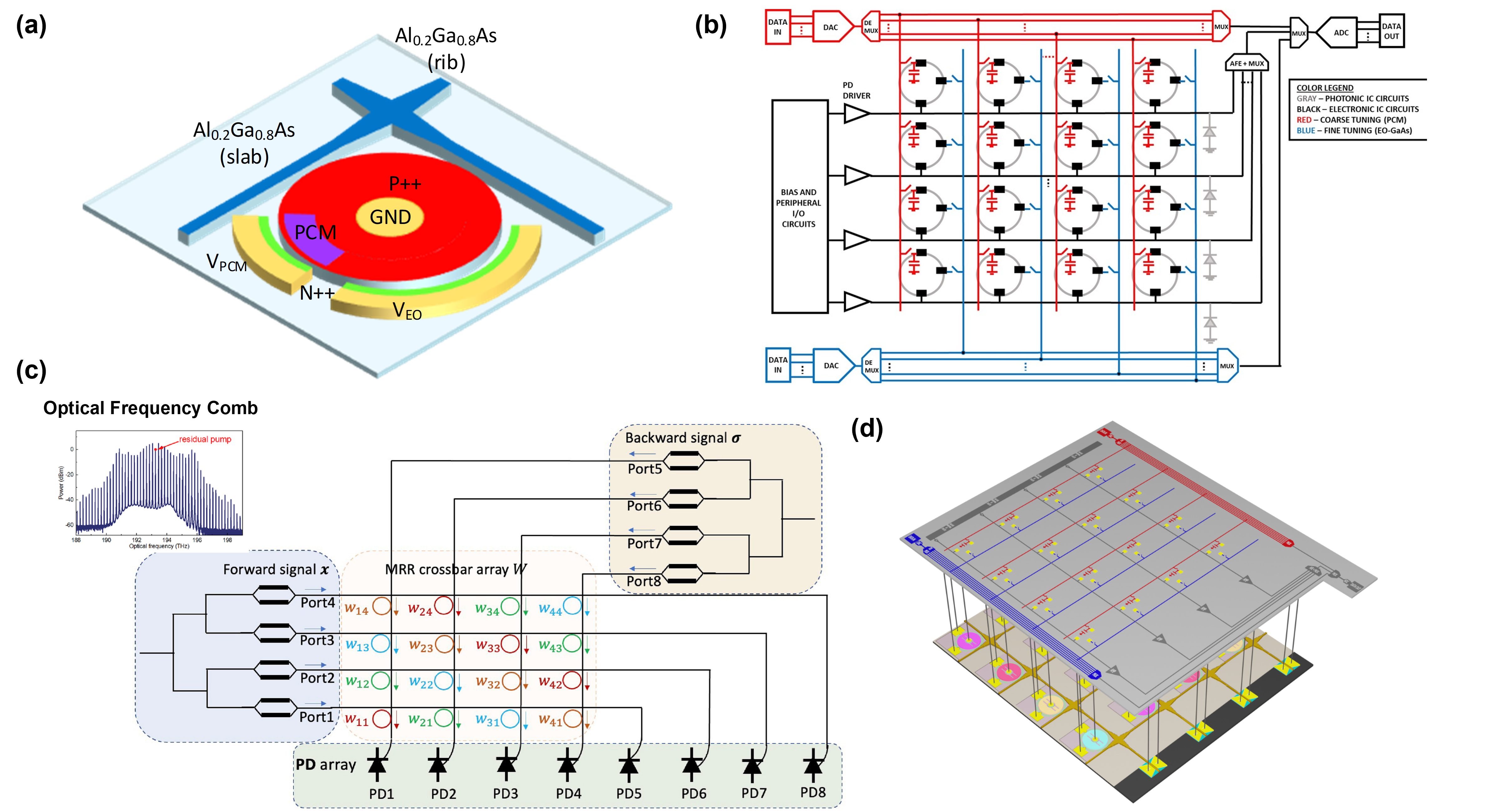}
		\caption{(a) PCM AlGaAs mem-resonator, composed of PCM (Sb2S3)~\cite{Rui:23} on AlGaAs-on-Insulator~\cite{Lin:19, Chang:18}, facilitates multi-precision tuning of weight values. (b) PCM-AlGaAs resonators are interconnected with a pulse circuit for PCM and run in parallel with capacitors (10pF) to create mem-resonators. These are driven by Digital-to-Analog Converters (DACs) arranged in a cross-bar configuration, transferring charges onto the PCM-AlGaAs mem-resonator to establish the desired voltage bias for the intended photonic weight matrix value. (c) In the data plane, the optical frequency comb (OFC) generating $>$32 combs will drive photonic tensors of size \textgreater 256\texttimes256. (d) The 3D integration of Electronic Integrated Circuits (EIC) and Photonic Integrated Circuits (PIC) through Direct Bond Interconnect (DBI\textsuperscript{®}) will realize the 3D-EPIC platform.}
		\label{fig:3D_system_architecture}
	\end{figure}

	\section{Hybrid 3D Mem-PCM Resonator PICs}
	The Photonic Integrated Circuit (PIC) comprises an array of hybrid memory resonators, combining Phase Change Materials (PCM) such as SbS or GST~\cite{Rui:23} with p-i-n AlGaAs ring resonators (see Fig.~\ref{fig:3D_system_architecture}(a)). As demonstrated in~\cite{Rui:23}, SbS can provide multiple levels (32 levels) of distinct non-volatile (NV) optical phase changes, enabling adjustments to weight values in the Photonic Integrated Circuits (PICs). The electro-optical effect of the p-i-n AlGaAs allows for a volatile phase shift in the AlGaAs ring by applying voltage across the p-i-n AlGaAs. Consequently, the PCM AlGaAs resonator achieves Most Significant Bit (MSB) NV phase tuning by applying pulses across the PCM, while the tuning of the Least Significant Bit (LSB) phase at low voltages can be accomplished by applying constantly reverse bias voltage.

	The p-i-n AlGaAs micro-disk resonator is 3D-integrated with Metal-Insulator-Metal (MIM) capacitors (with a total capacitance of approximately 10~pF on four Back-End-Of-Line (BEOL) metal layers) on an ultralow-leakage (\textless1pA) Fully-Depleted Silicon On Insulator (FD-SOI) CMOS Electronic Integrated Circuit (EIC) platform, such as GF22FDX. As illustrated in Fig.~\ref{fig:3D_system_architecture}(b), the crossbar electronic circuits on the CMOS EIC \cite{Wan:20,Kubendran:20,Wan:20_2,Wan:22} deliver an appropriate number of pulses to the PCM and an adequate amount of electrical charges into the hybrid capacitor (p-i-n AlGaAs and MIM capacitors in parallel). This precise control allows for the establishment of the desired voltage bias on the PCM AlGaAs mem-resonator for the specific wavelength, thus determining the intended photonic weight matrix value.

	The proposed photonic tensor core demonstrates a remarkable capability for frequent reprogramming, owing to its 1~pA leakage current and 10~pF capacitance. This combination ensures effective charge retention for approximately 100~ms, with thermal noise levels comfortably below 20~\textmu V.

	Consequently, the voltage retention across the PCM AlGaAs resonator achieves better than $1\times 10^{-4}$ accuracy over approximately 100~ms, assuming no additional leakage current is introduced. The flexibility of the system allows for reprogramming at a rate as high as the response time of the 14-bit DAC (about 10 \textmu s). Alternatively, the core can retain its program for extended periods, with a refresh cycle needed as infrequently as every 10~ms.

	The refresh cycle, designed to maintain stability, incorporates brief self-recalibration steps \cite{Miller:13,Miller:13_2,Miller:17,Ohno:22} to address any drift in bias voltages across the PCM AlGaAs mem-resonators. This comprehensive approach ensures the reliability and precision of the proposed photonic tensor core, making it a versatile and robust component for various applications.

	\section{PCM Materials}
	In optical  computing, PCMs can be employed in non-volatile memory.  These materials can undergo reversible phase changes based on optical or electrical stimuli. Following structural phase transitions from the covalent-bonded amorphous state to the resonant-bonded crystalline state, PCMs demonstrate significant variations in electrical resistivity and optical constants (typically $\Delta n > 1$) across a wide spectral range~\cite{Wuttig:07}. Once switched, the achieved state can endure for over ten years under ambient conditions without requiring any external power supply~\cite{Fang:21}.

	We aim to employ Ge\textsubscript{2}Sb\textsubscript{2}Te\textsubscript{5}~(GST) and Sb\textsubscript{2}S\textsubscript{3}~(SbS) for MSB programmability. Table~\ref{tab:pcm} summarizes the most common PCMs used in integrated photonics.
	\begin{table}[h!]
		\centering
		\caption{Comparison of refractive index change ($\Delta n$) and extinction coefficient ($kc$) from amorphous to crystalline state at wavelength of 1550~nm~\cite{Fang:21}.}
		\label{tab:pcm}
		\begin{tabular}{llll}
			\hline
				& $\Delta n$ & $k_c$ & $\Delta n/k_c$ \\ \hline
			GST    & 2.74       & 1.09  & 2.51           \\
			Sb2Se3 & 0.76       & 0     & Undefined      \\
			Sb2S3  & 0.54       & 0.05  & 10.8           \\ \hline
		\end{tabular}
	\end{table}
	While GST has high loss (1~dB) in its crystalline state at 1550nm, the wide bandgap of SbS with transparency windows ranging from 610 nm to near-IR allow large-scale PIC platforms and optical Field Programmable Gate Arrays (FPGAs)~\cite{Rui:23}. Therefore, GST becomes impractical for large-scale PIC platforms where light is guided through numerous phase change photonic routers.

	In~\cite{Zheng:20}, it is demonstrated the feasibility of inducing reversible large-area phase transitions over more than 1000 times (500 cycles) using low voltages (as low as 1 V for crystallization and 2.5 V for amorphization) by integrating GST on silicon PIN diode (p-type, intrinsic, n-type junction) heaters. Importantly, this process is achieved with near-zero additional loss. Another emerging low-loss PCMs such as Sb\textsubscript{2}Se\textsubscript{3}~\cite{Fang:21} may also be explored.

	PCMs offer excellent scalability and can be easily deposited on any substrate using sputtering, eliminating concerns about lattice mismatch. As a result, PCMs have found applications in compact, energy-efficient, and versatile programmable photonic integrated circuits (PICs) for switches, memories, and computing~\cite{Fang:21}.

	\section{3D Integrated System Architecture}
	As illustrated in Fig.~\ref{fig:3D_system_architecture}(b-d), the proposed
	PCM AlGaAs-OI platform comprises a hybrid PCM AlGaAs mem-resonator photonic-integrated circuit (PIC) that is 3D integrated~\cite{Zhang:20} with FD-SOI CMOS electronic integrated circuits (EIC). These electrical integrated circuits (EICs) serve as the programmable photonic tensor core~\cite{Bogaerts:20}. The system will be equipped with a low-noise, high-efficiency optical frequency comb (OFC) source~\cite{Liu:18}, additional periphery I/O, and control circuits (FPGA) with a user interface for the peripheral I/O circuitry.

	\subsection{PCM-AlGaAs Resonators and Unitcells }
	Various photonic technologies~\cite{Tossoun:24} exist for memory resonators; however, to date, there has been a lack of a non-volatile, low-loss memory resonators technology capable of achieving precise tuning of over 12-bits with repeatability, reliability, and speed. Recent advancements in PCMs have demonstrated low loss (0.01 dB) and multiresolution (5-bits), while AlGaAs materials have shown potential for low-loss, reliable, repeatable, and high-precision electro-optical tuning~\cite{morea:07,walker:91,eldada:94}.

    \begin{figure}[tb!]
        \centering
        \includegraphics[width=0.8\textwidth]{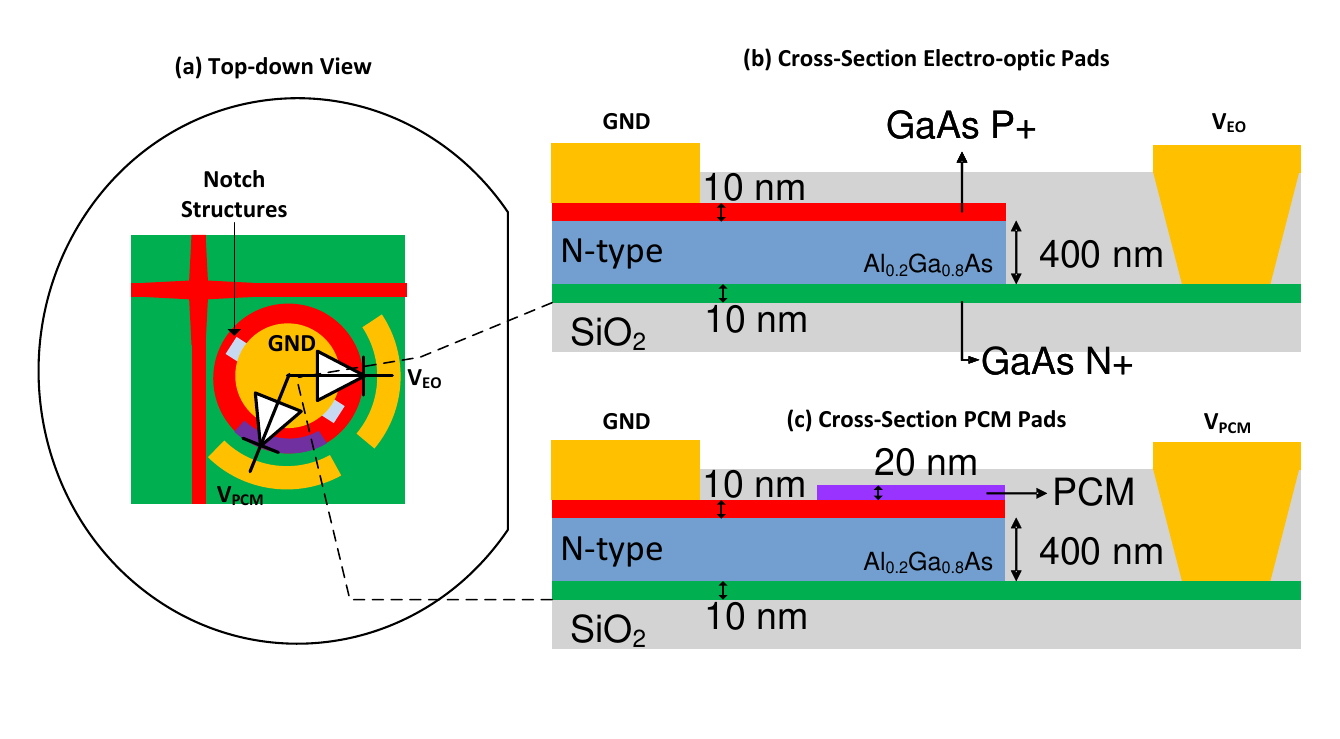}
        \caption{(a) Top-down view of the micro-disk resonator at the wafer level, and cross-section of the micro-disk modulator (b) across the electro-optic pads, and (c) PCM pads.}
        \label{fig:ring_topdown_view}
    \end{figure}

	This work aims to integrate these advancements to create the PCM AlGaAs memresonator. As illustrated in Fig.~\ref{fig:ring_topdown_view}, the PCM AlGaAs resonator comprises a vertically doped p-i-n structured AlGaAs micro-disk with a PCM material overcoat. The P+, and N+ layers are compromised by GaAs of thickness 10~nm, and doping concentration of $1\times 10^{18}$. Electro-optical tuning is achieved through reverse bias across the p-i-n structure (between V\textsubscript{EO} and GND) as shown in Fig.~\ref{fig:ring_topdown_view}(b), while pulsed heating for phase-changes in the PCM (amorphous $\Longleftrightarrow$  crystalline) is provided by a forward-pulsed voltage across the p-i-n structure (between Vpcm and GND), as depicted in Fig.~\ref{fig:ring_topdown_view}(c).

\begin{figure}[tb!]
            \centering
            \includegraphics[width=1\textwidth]{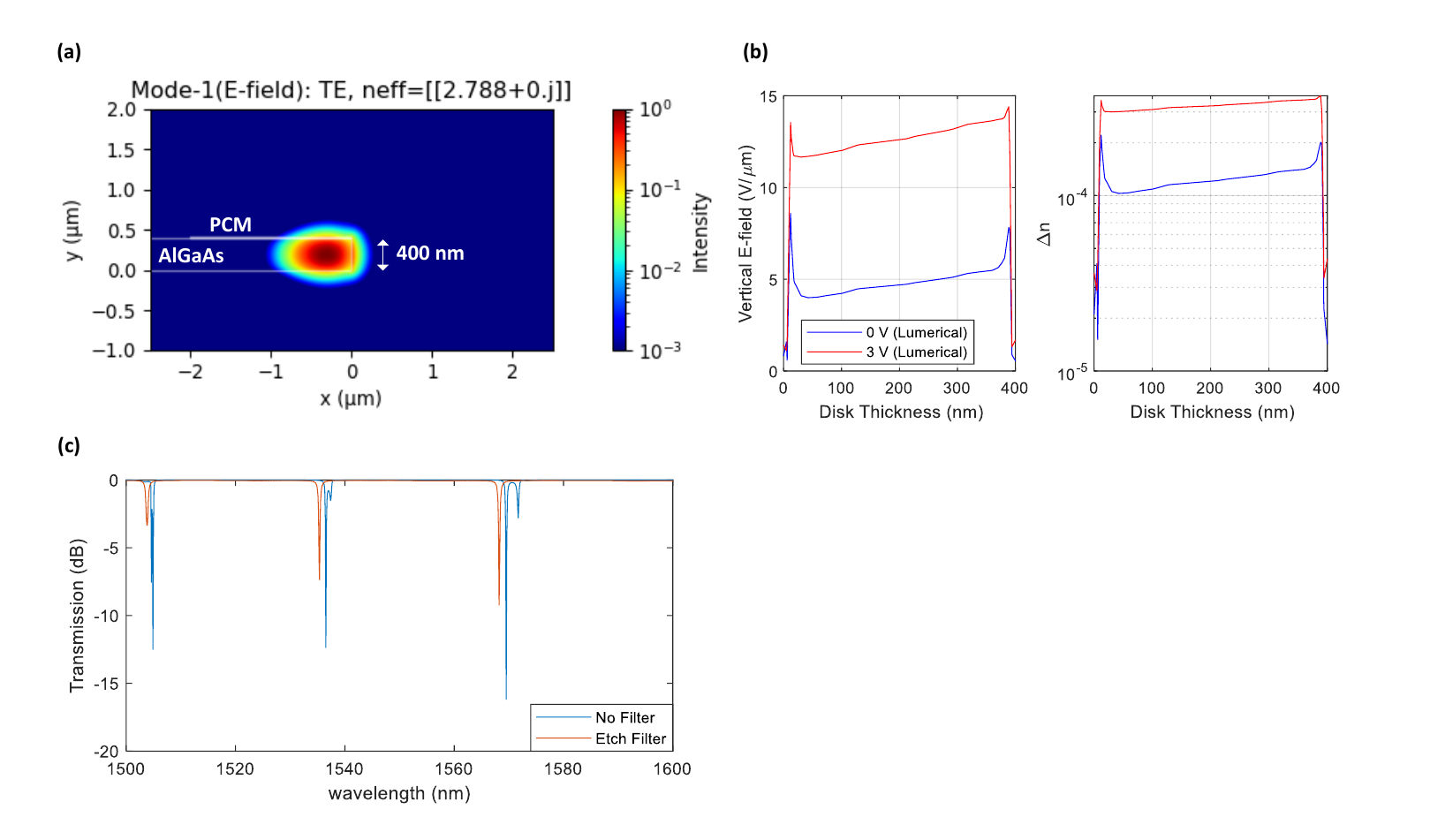}
            \caption{(a) Fundamental TE mode of the 400nm micro-disk waveguide overlaps with the 20nm PCM, (b) Magnitude of the E-field across the vertical p-i-n junction, and the index change across the horizontal direction of the micro-disk waveguide as a function of reverse bias voltage, (c) Frequency response of the disk modulator with and without the notch filter.}
            \label{fig:disk_results}
        \end{figure}

    Fig.~\ref{fig:disk_results}(a) illustrates the TE fundamental mode of the micro-disk waveguide, which has a thickness of 400~nm. The SbS layer on top of the micro-disk waveguide is 2~\textmu m in width and 20~nm in thickness. The confinement in SbS is approximately 1.5\%, with an effective index change from amorphous to crystalline of $\Delta n_{eff}=7\times 10^{-3}$, corresponding to a frequency change of about 130 GHz in the resonant wavelength of the micro-disk.  The E-field magnitude across the vertical p-i-n junction starts to increase as the reverse bias increase as depicted in Fig.~\ref{fig:disk_results}(b). As the magnitude of the E-field across the thickness of the micro-disk waveguide starts to increase, a change of the index in the lateral direction in AlGaAs micro-disk wavegude is occurred. Leveraging the electro-optical tensor's modulation of the TE mode optical index due to vertical electrical bias, the TE mode optical resonance is utilized. An index change at the order of $\Delta n_{eff}=2\times 10^{-4}$, will induce an electro-optical bandwidth of about 4~GHz. Notch structures etched into the micro-disk serve to suppress high-order modes, ensuring single-mode operation as illustrated in Fig.~\ref{fig:disk_results}(c). Experimental results with a similar ring structure AlGaAsOI have demonstrated a intrinsic Q-factor of 1.5~\texttimes~10\textsuperscript{6}, which corresponds to a propagation loss around 0.4 dB~cm\textsuperscript{-1}~\cite{Lin:19}.
 
\begin{figure}[tb!]
		\centering
        \includegraphics[width=0.9\textwidth]{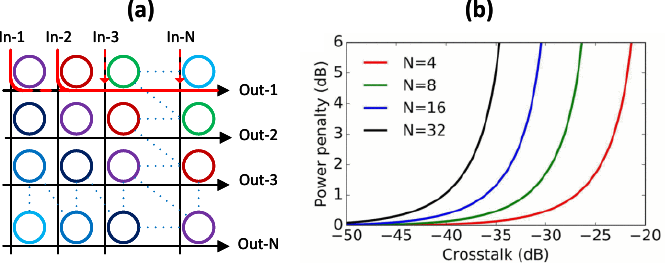}
		\caption{(a) Crossbar Ring Resonator array, and (b) Crosstalk penalty of AlGaAs ring  resonator~\cite{Xiao:17}}
		\label{fig:napsac_ring_crosstalk}
	\end{figure}

     In the context of W~\texttimes~W photonic tensors, the optical crosstalk can incur a penalty that increases superlinearly with the parameter W~\cite{Xiao:17}, as illustrated in Fig.~\ref{fig:napsac_ring_crosstalk}. The underestimation of this penalty was noted in earlier publications~\cite{Bianco:10}, and we addressed and rectified it through the introduction of~(1) in~\cite{Xiao:17}.

	For the mem-resonator design depicted in Fig.~\ref{fig:3D_system_architecture}(a), the input signal can resonantly drop (shown as 'On') or to go through (shown as 'Off'). For on-state (off-state), the incident light outputs from the drop-port (the through-port) with an insertion-loss $IL_{\text{on}}$ ($IL_{\text{off}}$) with crosstalk $X_{On}$ ($X_{Off}$).  We obtain an intra-band signal-crosstalk beat-noise of $\sigma_{RIN} = 8.98 \times 10^{-8}$ and $\text{SNR}_{RIN} = 70.46~\text{dB}$, $IL_{\text{on}}$ ($IL_{\text{off}}$) $IL_{\text{on}} = 1.13 \ \text{dB}$, $IL_{\text{off}} = 0.03 \ \text{dB}$, and cross-talk $X_{\text{on}} = -18.23 \ \text{dB}$, and $X_{\text{off}} = -70 \ \text{dB}$ at 200 GHz. The crosstalk remains below -80 dB after 400 GHz, supporting an Effective Number of Bits (ENOB) of 12-bits in the optical 32x32 crossbar.

	The photonic tensor core's unit cell comprises a waveguide crossing, which will manifest both loss and crosstalk. The resultant Relative Intensity Noise (RIN) calculated using equation (1) in~\cite{Xiao:17} drops below -160~dBc/Hz for W=32.

	\subsection{Flatband 1\texttimes 4 interleavers and 4\texttimes 1 de-interleavers}
	An optical interleaver is a device used in optical systems to separate and combine multiple wavelengths of light. For example, it is commonly used in wavelength-division multiplexing (WDM) systems, where different wavelengths of light are transmitted simultaneously over a single optical waveguide. We primarily focus on a ring-assisted Mach-Zehnder Interferometer (RAMZI); the MZI consists of two 3-dB directional couplers, each stage with two ring-resonators. The phase in each arm is constant and tuned based on a given 3-dB bandwidth filtering function. Microheaters adjust these phase values to tune the resonance frequency of the rings and compensate any fabrication imperfections.

\begin{figure}[tb!]
		\centering
		\subfloat[]{\includegraphics[width = 0.4\textwidth]{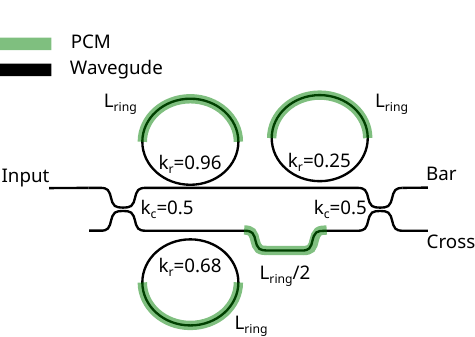}
			\label{optical_interleaver_1st}}
		\subfloat[]{\includegraphics[width=0.45\textwidth]{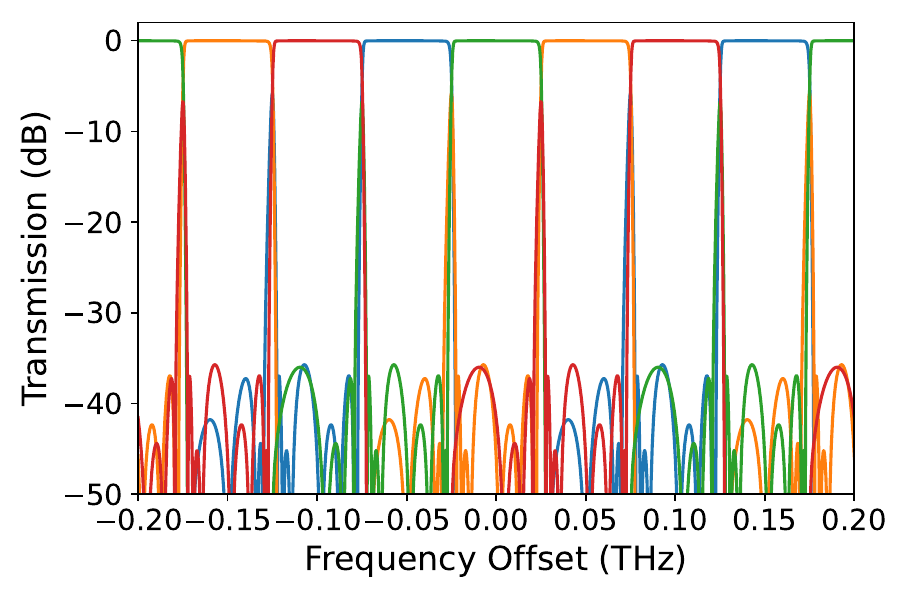}%
			\label{optical_interleaver_2nd}}
		\caption{50~GHz (de-)interleaver (a) Scheamatic diagram and (b) transmission response with phase change material (PCM) of two-ring RAMZI.}
		\label{fig:optical_interleaver}
	\end{figure}

	The hierarchic scaling of the $N\times N$ system will incorporate $L_o\times 1$ wavelength interleaving. The current design will incorporate $W=32$ wavelengths and $L_o=4$ interleaved stages for $N=128$. The frequency comb generator shown in Fig.~\ref{fig:3D_system_architecture}(c) will incorporate $N$~lines at 10~GHz spacing. The interleaver at the detector will allow the summing of the optical power at each comb line such that the crosstalk rejection of the interleavers become less critical. Fig.~\ref{fig:optical_interleaver} shows the interleaver for a channel spacing of 50~GHz, and an FSR at 200~GHz employs trimming capability by PCM  to achieve non-violate trimming of the rings and couplers.

	\section{Low-Noise, Ultra-High Efficiency, Photonic-Crystal OFC}
	Low-noise, high-power efficiency in optical frequency comb (OFC) generation is extremely important for the proposed system. We will investigate Kerr nonlinear microresonators to convert a continuous wave (CW) pump laser into a "microcomb." Soliton microcombs offer a high repetition frequency and a very broadband output, supporting hyper-parallelization at hundreds of optical channels.We will design, fabricate, and test the soliton microcomb, showcasing its capabilities in demonstrations alongside the full in-memory computation system. In particular, we will leverage photonic-crystal resonator (PhCR) solitons which have emerged from \cite{yu:22,yu:21}.

\begin{figure}[tb!]
		\centering
		\includegraphics[width=0.9\textwidth]{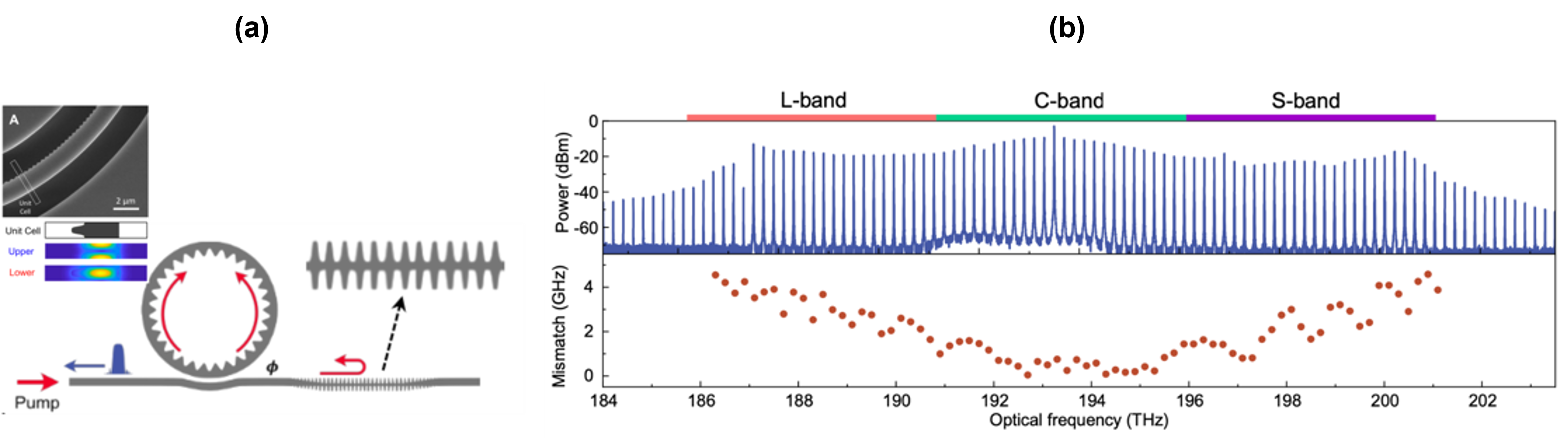}
		\caption{(a) Photonic crystal resonators for comb generation. (b) Spectrum covering the entire SCL telecom band with ITU grid alignment.}
  		\label{fig:comb_generator}
	\end{figure}

	As depicted in Fig.~\ref{fig:comb_generator}, PhCRs are resonant structures where a nanopattern is etched onto the inner edge of a ring resonator, creating a photonic bandgap within the resonator's mode structure. PhCR solitons enable the highly efficient generation of solitons with repetition frequencies ranging from less than 50 GHz to hundreds of GHz or even THz, utilizing a 1550~nm pump-laser source.

	The specific requirement for this platform is the development of PhCR soliton generators featuring a 200~GHz mode spacing, 32~channels, and low intensity and phase noise of the soliton microcomb. Four separate microcomb units will be created, each with a programmed offset frequency of 50~GHz, resulting in a total of 128~channels. The middle panel of Fig.~\ref{fig:comb_generator} illustrates the precise spectrum control essential for this task.

	Solitons in Kerr resonators represent isolated nonlinear eigenstates of the intraresonator field, influenced solely by the material properties of the resonator, dissipation, the pump laser, and quantum fluctuations. Photonic bandgaps within  PhCRs introduce mode-specific frequency shifts, facilitating microcomb generation in either bright soliton or dark soliton modality. This offers the flexibility of tailoring soliton spectra to meet specific application requirements.

\begin{figure}[tb!]
		\centering
		\includegraphics[width=0.9\textwidth]{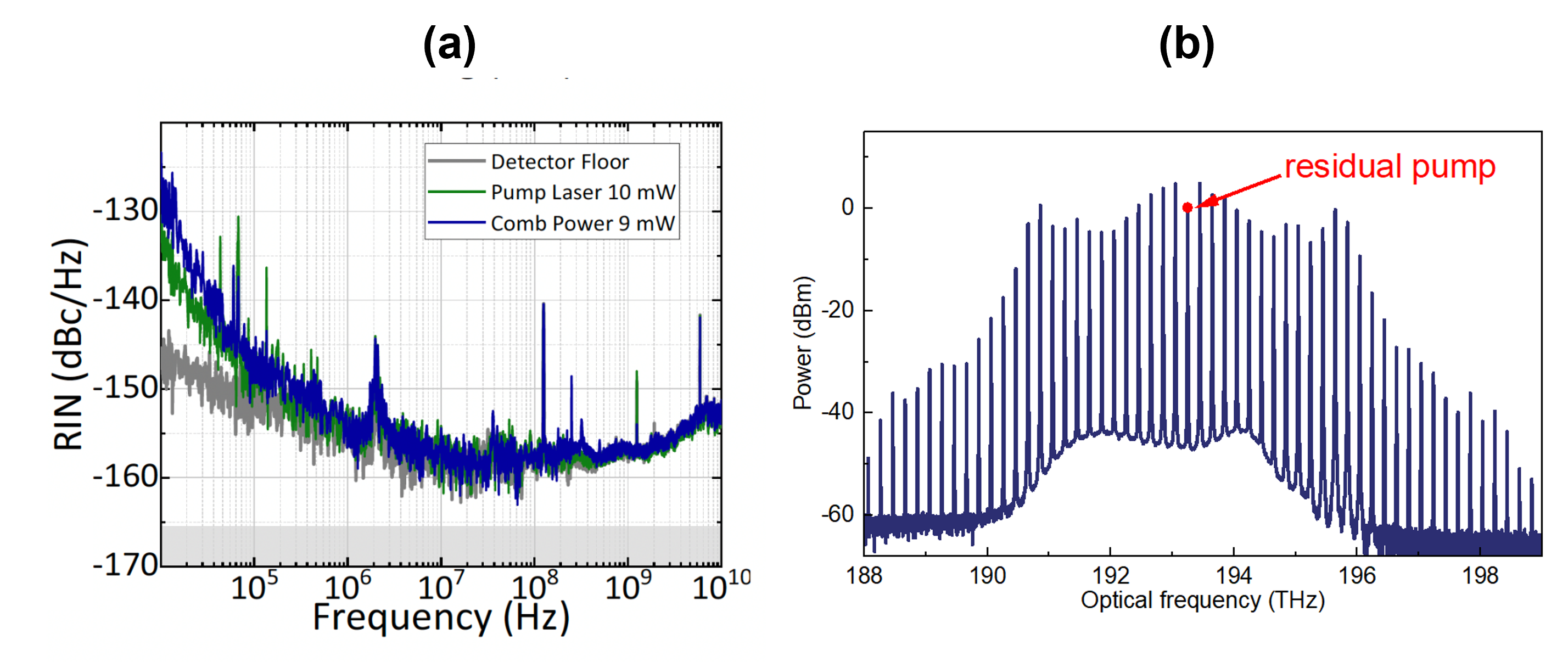}
		\caption{(a) 200 GHz comb with suppressed pump and ~32 comb lines at ~ 0 dBm.  (b) RIN measurements of the comb indicating -160 dBc/Hz RIN performance limited by the detector noise floor.}
  		\label{fig:comb_rin}
	\end{figure}

	The frequency-shifted mode can be engineered to enable four-wave mixing with a higher capability compared to an un-patterned resonator. Dark-soliton microcombs provide unprecedented continuous-wave laser wavelength conversion, as illustrated in the left panel of Fig.~\ref{fig:comb_rin}, where the residual pump power in a PhCR is lower than the nearby comb lines. Our findings indicate consistently high device conversion efficiencies (\textgreater50\%) with tens of milliwatts of on-chip pump power, showcasing the robustness of the fabrication.

	We further aim to investigate into the device dynamics that aims to enhance conversion efficiency for high-power hyperspectral PhCR microcomb sources compatible with the PCM AlGaAs-MemResonator in-memory computing system. As depicted in Fig.\ref{fig:comb_generator}, the achievement of 32~comb lines at 0~dBm, with an optical power conversion efficiency exceeding 65\% and a Relative Intensity Noise (RIN) below -150~dBc/Hz, underscores the readiness of this technology for seamless integration into the system.

	\section{Electronic IC Technologies}
	The Electronic Integrated Circuit (EIC) will encompass the design and implementation of low-power, high-speed, and high-precision electronic circuits. These circuits will subsequently be integrated with photonic-integrated-circuits (PIC). The EIC design incorporates data converters, analog front-end circuits, low-leakage switches, calibration, and peripheral circuitry to control and tune the mixed-precision PCM AlGaAs mem-resonators.

	To achieve precise and independent tuning of the PCM device and AlGaAs on insulator ring resonator, the PCM-AlGaAs resonators are connected with a pulse circuit for PCM. Simultaneously, they are connected in parallel with capacitors (10pF) to form mem-resonators driven by Digital-to-Analog Converters (DACs) configured in a cross-bar array. This array transfers charges onto the PCM-AlGaAs mem-resonator to set the desired voltage bias for the intended photonic weight matrix value.
	
	The EIC will be designed as a module for a 32~\texttimes~32 photonic tensor core and can be scaled to a 256~\texttimes~256 crossbar array or even further. Each EIC module will consist of 32 TIA (Trans-Impedance Amplifier) stages for reading photodetector outputs from the PIC. It will also include four 6-bit DACs and four 10-bit DACs with a sampling rate of 200-MSPS to tune the PCM and Electro-optic AlGaAs, respectively. Additionally, four 12-bit ADCs with a sampling rate of 500-MSPS will be included to read out the PCM/AlGaAs element conductances and the photoreceptor outputs. The EIC will also feature I/O drivers, bias circuits, and other peripheral circuits to program the EIC components, including DACs, ADCs, and TIA stages.

    Optical photodetectors in the PIC will utilize photodiodes to convert the light intensity to photocurrents, which will be read out and provided as input to the EIC. Up to 16 photodiodes will connect to each photodetector Analog-Front-End (AFE) using interposers for space division multiplexing or interleavers for wavelength division multiplexing. Each AFE will consist of a high gain (\textgreater90dB), low power (\textless10\textmu W), and low noise (\textless10\textmu V/Hz) trans-impedance amplifier (TIA) followed by a bandpass filter (BPF) that provides the amplified and filtered output voltage to the ADC for digital readout.

    Several separate arrays of DACs will be used in each EIC module for tuning the conductance of the PCM and electro-optic AlGaAs mem-resonator elements in the PIC. For coarse tuning of 32\texttimes32 PCM conductances, a low-resolution DAC is sufficient, such as a 6-bit DAC. Four such DACs will be used to tune a 32\texttimes32 crossbar array, implying one DAC will tune 256 PCM elements using time-division multiplexing. These DACs will be implemented as thermometer DACs with a unit cell providing 1~LSB. For fine-tuning of 32\texttimes32 electro-optic AlGaAs mem-resonator elements, a high-resolution DAC is needed, such as a 10-bit DAC. Four such DACs will be used to tune a 32\texttimes32 crossbar array, implying one DAC will tune 256 AlGaAs mem-resonator elements using time-division multiplexing. These DACs will be implemented as segmented DACs with 4 LSB binary coding, and the remaining MSB (4-6 bits) will be using thermometer coding. All DACs will be designed at a sampling rate of 100-200 MSPS to provide sufficient high-speed configuration of the output voltage to tune 256 elements within a short time of 10~ms, which maps to 40~\textmu s for each element. The ENOB of the combined mixed-precision system will be designed to achieve \textgreater8 bits ENOB.

    Four high-resolution, high-speed sigma-delta ADCs will be implemented for each EIC module, addressing a 32\texttimes32 photonic tensor core. Initially, the ADC will target 10-bit resolution (ENOB \textgreater 8 bits) at a 300-MSPS sampling rate. Later, the ADC will be upgraded to 14-bit resolution (ENOB \textgreater 12 bits) at a 500-MSPS sampling rate. The ADCs can be used for two modes. (a) Reading the optical photodetector array output providing the AFE stage, and (b) Reading the conductance of the PCM and AlGaAs mem-resonator elements. Each ADC will address 256 elements using time-division multiplexing to read the photodetector AFE output or the conductance of the mem-resonator elements using a TIA-based integrator output.
    
	\section{Fabrication, microtransfer-printing, and 3D Integration}
\begin{figure}[tb!]
		\centering
		\includegraphics[width=0.9\textwidth]{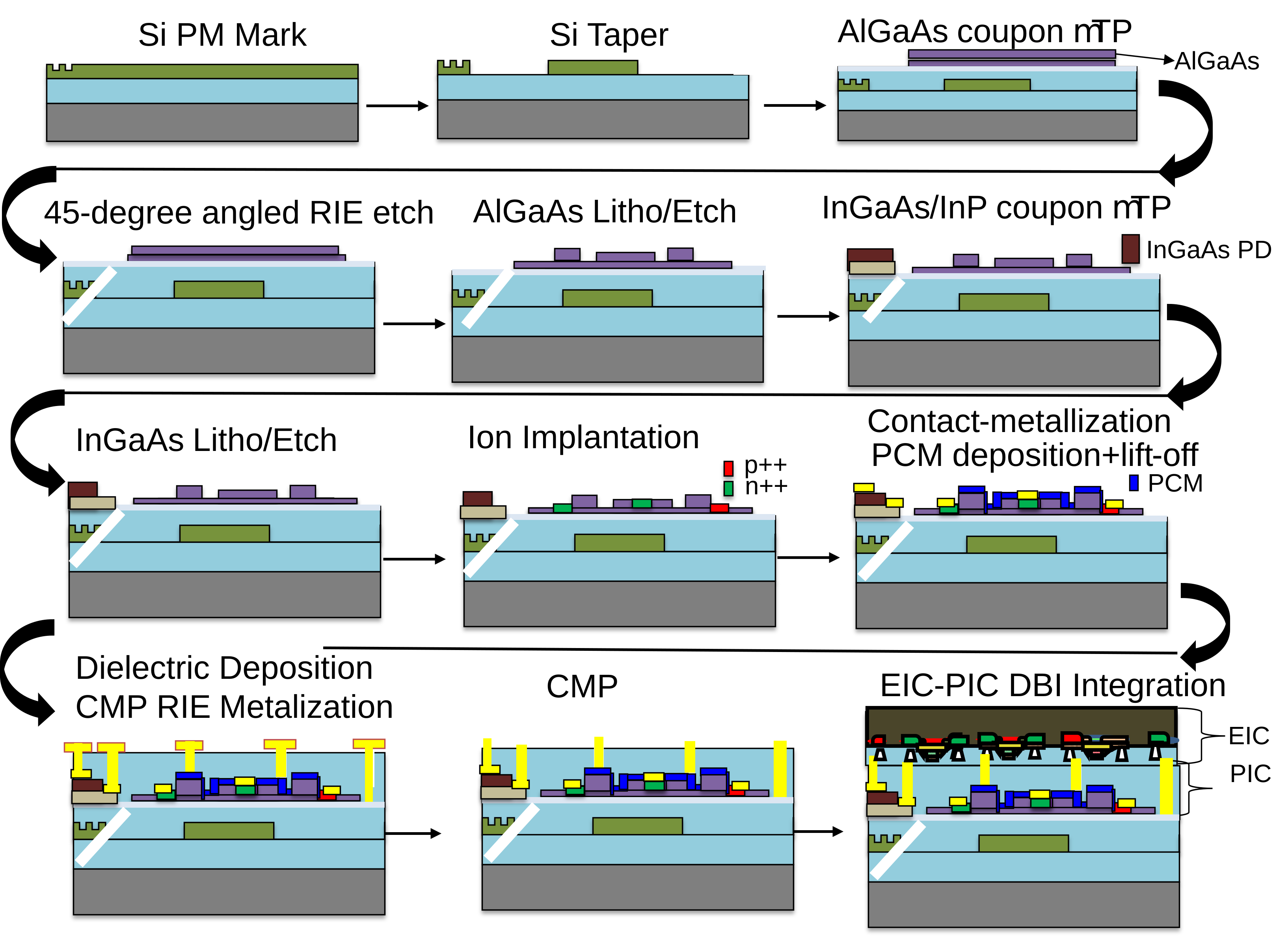}
		\caption{Fabrication and heterogeneous integration steps for 3D EPIC tensor core}
		\label{fig:fabrication_chart}
	\end{figure}

	The proposed system leverages advanced heterogeneous integration, 150-nm resolution CMOS fabrication augmented by 10-nm resolution e-beam lithography, micro transfer-printing (µTP), and 3D integration. We 	will manufacture the photonic tensors on a 150 mm wafer scale using the ASML stepper and µTP, as depicted in Fig.~\ref{fig:fabrication_chart}. For InGaAs photodetectors (PDs), the process involves 45-degree angled etching and InGaAs µTP. For foundry wafers with Ge detectors, these steps can be omitted. The state-of-the-art 3D EIC-PIC integration through direct bond interconnect (DBI\textsuperscript{®}) \cite{Jiang:17,Agrawal:17,Samantha:23} represents an advanced 3D integration solution, by merging the top metal and dielectric of two wafer/die, offering a bond pitch as small as 2~$\mu$m.

\begin{figure}[tb!]
		\centering
		\includegraphics[width=0.9\textwidth]{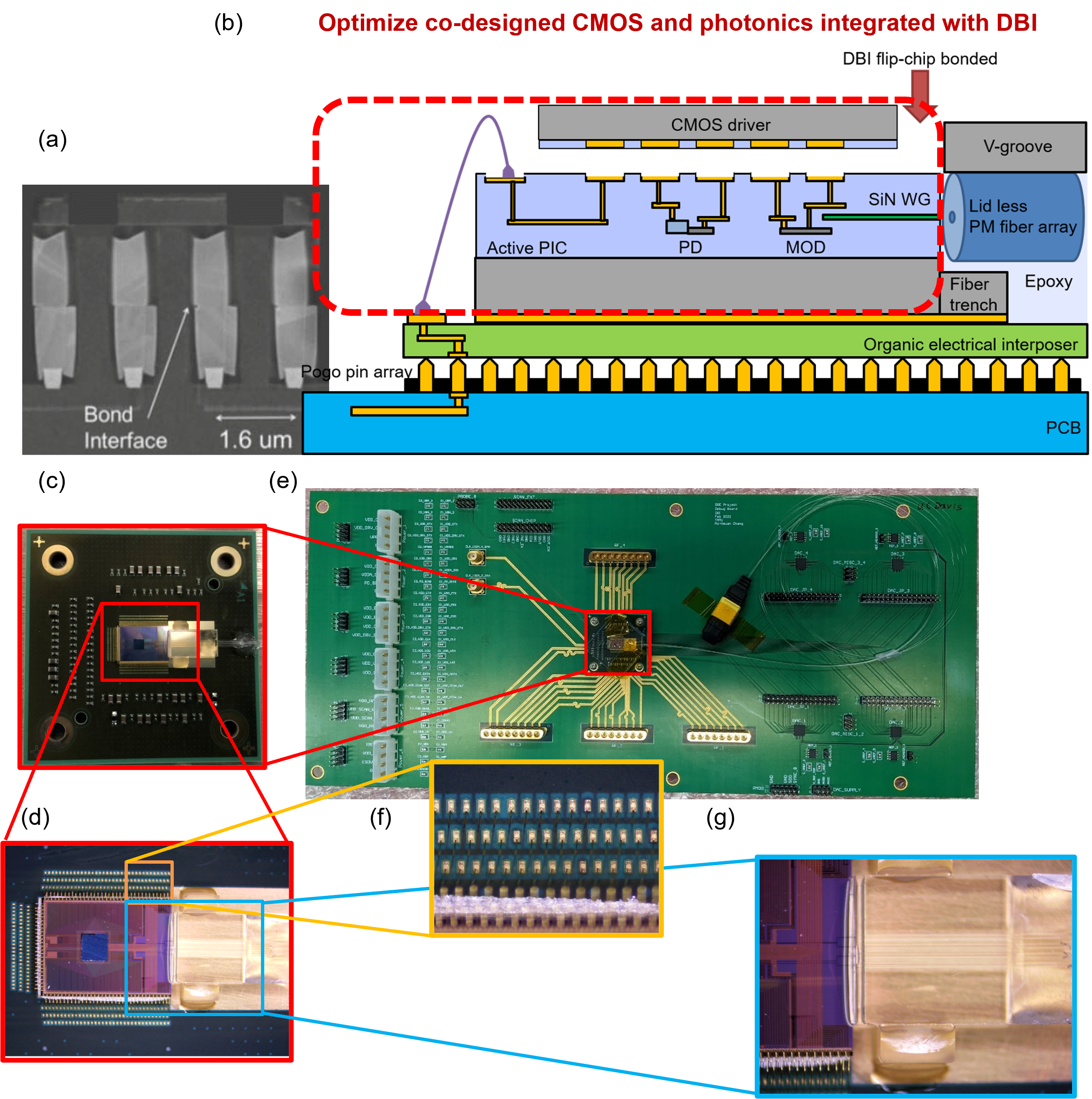}
		\caption{Photo (a) displays the DBI\textsuperscript{®} bonding cross-section, while (b) presents a schematic of the 3D EPIC on interposer. Photos (c-g) showcase the final packaged transceiver module with a fiber array, featuring a 12nm FinFET Electronic Integrated Circuit.}
		\label{fig:dbi_bonding}
	\end{figure}

	Crucial to our objectives is the establishment of efficient interconnections with minimal parasitic effects between photodetectors and CMOS transimpedance amplifiers (TIAs), resulting in a noteworthy 6~dB enhancement in receiver sensitivity. A recent breakthrough at UC Davis showcased photoreceiver arrays achieving record-high sensitivity~\cite{Chang:23}. This accomplishment involved the integration of 12~nm CMOS electronic circuits with silicon photonics 32-channel receivers and transmitters, operating at a remarkable efficiency of 496~fJ/b at 25~Gb/s. The 3D EPIC silicon photonic photonic integrated circuit (PIC) transceivers, illustrated in Fig.~\ref{fig:dbi_bonding}, are seamlessly adaptable to FPGA interfaces, facilitating the creation of peripheral I/Os essential for the computing platform.

	\section{Photonic Tensor Core with Modular Scaling in wavelength-space domains}
\begin{figure*}[tb!]
		\centering
		\includegraphics[width=1.0\textwidth]{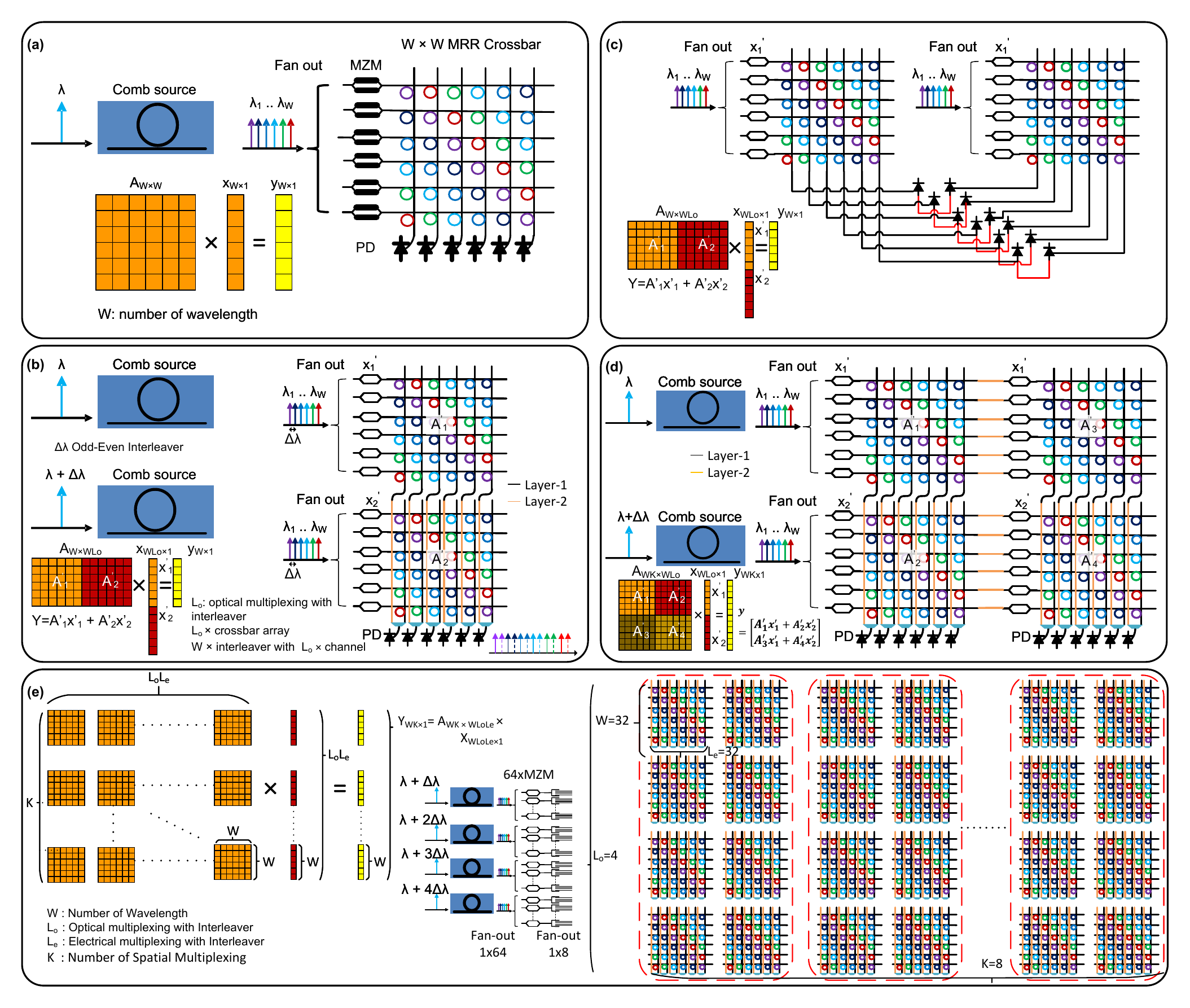}
		\caption{Scaling options:
		(a) Scaling of a W\texttimes W photonic tensor core pumped by one OFC source with $W$ wavelengths to $W^2$ resonators. (b) Scaling of LoW\texttimes W by utilizing a low number of OFC sources interleaved in wavelength domains. (c) Scaling of LeW\texttimes W by utilizing a low number of spatial division multiplexing. (d) Scaling of WLo\texttimes WLo by again utilizing a low number of OFC sources interleaved in a low number of wavelength domains. Here, W can be 32 based on the conservative estimate of crosstalk mitigation limit, and Lo can be 4 based on the practical limit.(e) KWxLoLeW photonic tensor core pumped by Lo number of OFC source of W wavelengths.Lo can be 4, Le can be 2, and K=LoLe=8 so that system can complete 256x256 tensor for 256x1 vector solutions.}
		\label{fig:wxw_tensor}
	\end{figure*}

	We aim to create an innovative, modularly scalable hyperdimensional photonic computing architecture featuring a photonic tensor with a size exceeding 256~\texttimes~256, supporting a precision of 12-bits. Figs.~\ref{fig:wxw_tensor} illustrate the modular scaling path to achieve greater than 1~\texttimes~128, utilizing a 32~\texttimes~32 crossbar. Additionally, a dual 256~\texttimes~256 crossbar configuration, as shown in Fig.~\ref{fig:wxw_tensor}(e), will be implemented in twin setups, both pumped by the same four OFC sources.

	Further scalability can be achieved by incorporating multiple Free Spectral Ranges (FSR). We have previously successfully designed a Photonic Tensor Core of size 1024~\texttimes~1024 using Tensor Train (TT) Decomposition methods, integrating multiple wavelengths~\cite{Xian:21}. This core comprises multiples of 8~\texttimes~8 TT cores, resulting in 582 times fewer photonic components compared to fully-connected 1024~\texttimes~1024 photonic meshes (with approximately 1 million elements). Remarkably, this design maintains negligible reduction in accuracy.

	\section{Noise, loss, crosstalk, and System ENOB}
\begin{figure}[tb!]
		\centering
        \includegraphics[width=1.0\textwidth]{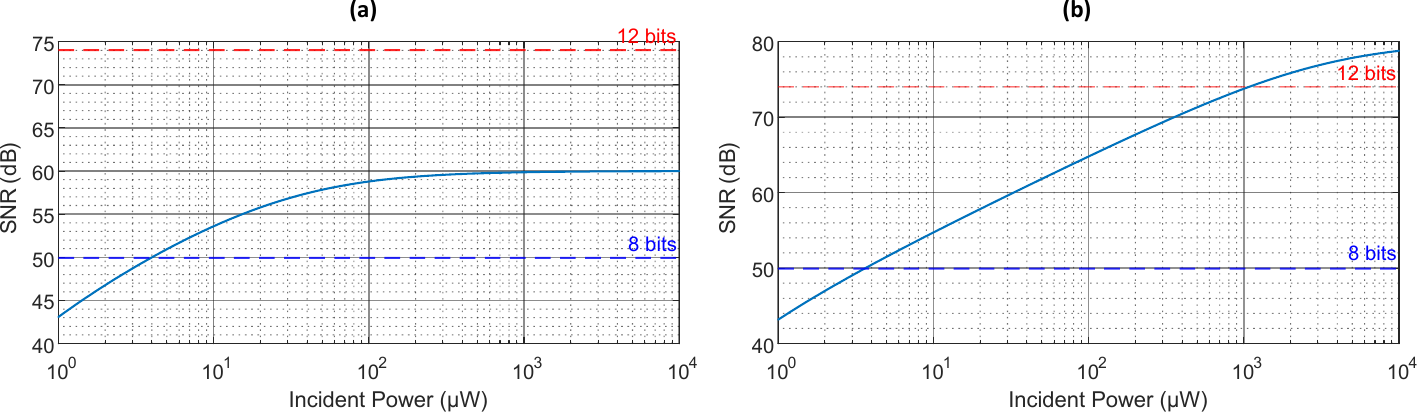}
		\caption{Requirements at the p-i-n detector based system to achieve ENOB~\textgreater~8-bits, and ENOB~\textgreater~12-bits for a RIN (a) -140~dBc/Hz, and (b) -160~dBc/Hz per comb-line. }
		\label{fig:ENOB}
	\end{figure}

	We have conducted Effective Number of Bits (ENOB) calculations, encompassing various noise sources such as photodetectors, optical source Relative Intensity Noise (RIN), jitter, microresonator crosstalk, electrical circuit noise, and others. Fig.~\ref{fig:ENOB} summarizes the results, indicating that an ENOB of 12-bit is achievable with approximately 1 mW of incident power on the photodetector for a 100 MHz bandwidth, while higher power is necessary at higher speeds. Additionally, it necessitates the use of an optical source with an (individual combline) RIN of less than -160 dBc/Hz.

	\section{Applying scientific computing PDEs to Photonic Tensors with Mixed Precisions}
	Numerical algorithms for PDEs, describing applications such as flow, transport, and mechanical response in this proposal, can be converted into tensor-vector multiplications. These conversions can then be implemented in the proposed engine through pipelined or recursively looped back modules (see, e.g.~{\cite{Bogaerts:20}}).

	The proposal will incorporate the mixed-precision in-memory computing algorithm~\cite{Le_Gallo:18} into the mixed-precision PCM-AlGaAs memory resonators hardware itself to achieve high precision. Additionally, the in-memory error detection method in~\cite{Ohno:22} will be integrated into the on-chip balanced detection system to achieve iterative refinement without relying on DRAMs.

\begin{figure}[tb!]
		\centering
		\includegraphics[width=0.9\textwidth]{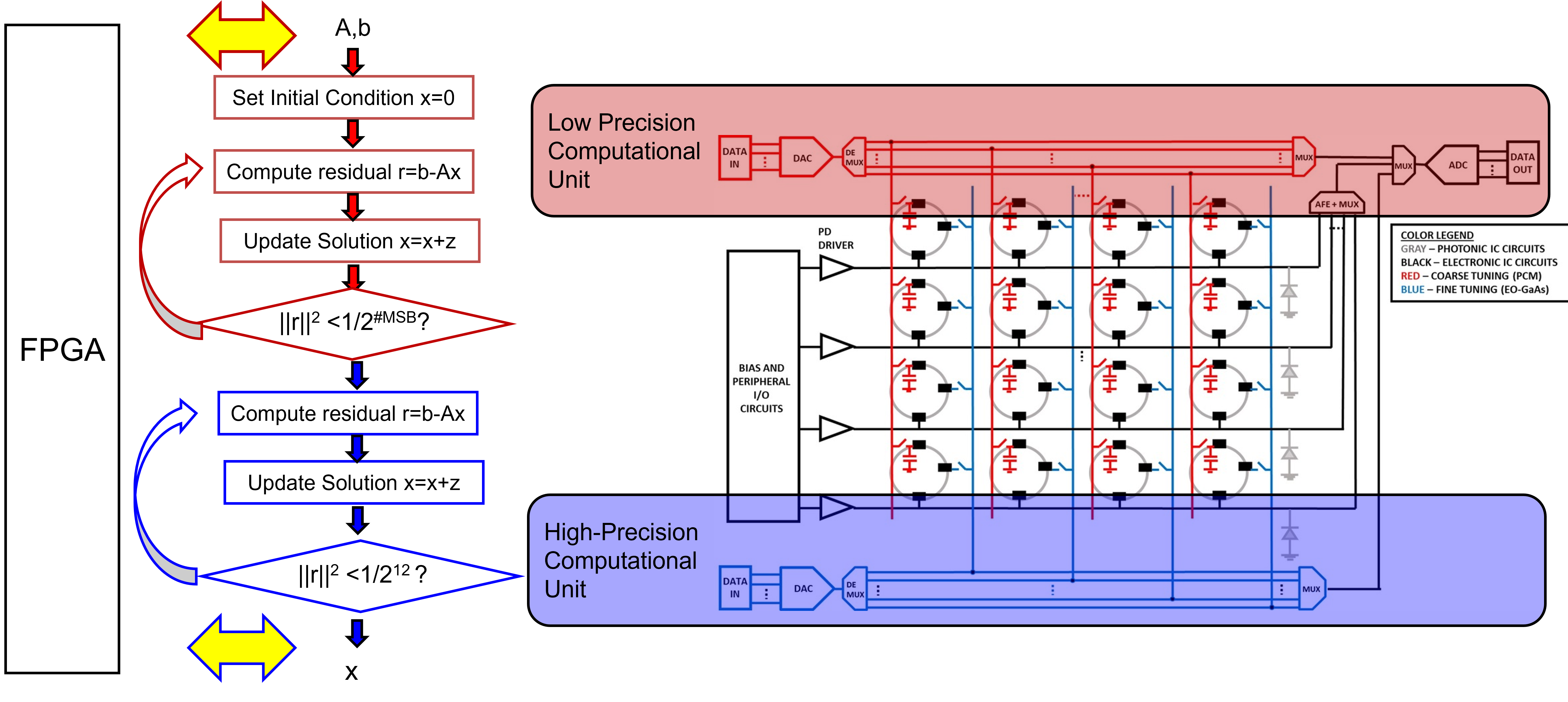}
		\caption{Concept of mixed-precision in-memory computing: a Possible architecture of a mixed-precision in-memory computing system. The FPGA trains the matrix A following the method of~\cite{Ohno:22}, and the low-precision computational memory unit (blue) performs analogue in-memory computation using the PCM-AlGaAs memresonator arrays by tuning the PCM. The in-memory computation with balanced detection calculates the residual r=b-Ax. This process iterates with PCM until the error becomes smaller than the precision of PCM (MSB=5 in this case). Then, the high-precision computational memory unit (red) performs analogue in-memory computation using the PCM-AlGaAs memresonator arrays by electro-optical tuning p-i-n AlGaAs micro-disk. Again, the in-memory computation with balanced detection calculates the residual r=b-Ax.}
		\label{fig:algorithm}
	\end{figure}

	By modifying Fig.~\ref{fig:3D_system_architecture} of ~\cite{Le_Gallo:18}, we obtain Fig.~\ref{fig:algorithm}, where the high-precision electro-optical part of the PCM-AlGaAs memresonator replaces the Von-Neumann computing in~\cite{Le_Gallo:18} using the algorithm described in the caption. The residual in the iterative linear equation system solver and the updated solution can leverage balanced detection and analog current sum by modifying the method described in~\cite{Ohno:22}, without relying on memory and processors. The analog computation result can directly adjust the voltages of the modulators to update the solution.

	In Fig.~\ref{fig:algorithm}, the FPGA will handle control and initial/final I/O but will not require Von Neumann computing. The mapping of large-scale scientific applications to multiples of photonic tensors with reconfigurability is deemed extremely important. Referring to the TT core in Fig.~\cite{Xian:21}, we demonstrated an architecture for 1024 x 1024 tensor computation using 32 wavelengths and 8x8 tensor cores with 582 times fewer components.

	\section{Conclusion}
	The envisioned project is dedicated to the realization of photonic in-memory computing through the integration of 3D-Photonic-Electronic circuits, incorporating Phase-Change-Material (PCM), AlGaAs, and CMOS technologies. The primary goals include achieving an exceptional level of accuracy surpassing 12-bits, ensuring high scalability exceeding 1024 by 1024 array dimensions, and implementing extreme parallelism within the Wavelength-Space-Time domains, surpassing a remarkable 1 million parallel processes. All of this is to be achieved at an ultra-low power consumption of less than 1~Watt per PetaOPS.
	
	The proposed architecture will involve the comprehensive development, validation, and bench-marking of a groundbreaking modality of scalable, ultra-low power 'In-memory' computation. This novel approach is characterized by its exceptionally low Size, Weight, and Power (SWaP) requirements, promising high throughput, and adaptive programmability. The anticipated outcomes of this research hold the potential to revolutionize computing paradigms, offering a versatile solution applicable across a wide spectrum of applications. The focus lies not only on pushing the boundaries of computational accuracy and scalability but also on ensuring efficiency and adaptability in real-world scenarios. Through this innovative approach, the project aims to usher in a new era of computing that aligns with the demands of various applications while operating at the forefront of technological advancements.


\bibliography{sn-bibliography}

\end{document}